\documentclass[10pt,letterpaper]{article}
\usepackage{opex3}
\usepackage[dvips]{epsfig}
\usepackage{float}
\usepackage{amsmath}
\usepackage{amssymb}
\usepackage{pslatex}
\usepackage{hhline}
\usepackage{threeparttable}

\newcommand{\vrr}{\mathbf{r}}

\begin{document}

\bibliographystyle{osa}

\title{Hybrid photonic crystal cavity and waveguide for coupling to diamond NV-centers}

\author{Paul E. Barclay,  Kai-Mei Fu, Charles Santori and \\ Raymond G. Beausoleil}
\address{Hewlett-Packard Laboratories, 1501 Page Mill Road, Palo Alto CA 94304}
\email{phone: (650) 857-6119, e-mail: paul.barclay@hp.com}

\begin{abstract}
A design for an ultra-high $Q$ photonic crystal nanocavity engineered to interact with nitrogen-vacancy (NV) centers located near the surface of
a single crystal diamond sample is presented. The structure is based upon a nanowire photonic crystal geometry, and consists of a patterned high
refractive index thin film, such as gallium phosphide (GaP), supported by a diamond substrate. The nanocavity supports a mode with quality
factor $Q > 1.5 \times 10^6$ and mode volume $V < 0.52 (\lambda/n_\text{GaP})^3$, and promises to allow Purcell enhanced collection of
spontaneous emission from an NV located more than 50 nm below the diamond surface.  The nanowire photonic crystal waveguide can be used to
efficiently couple light into and out of the cavity, or as an efficient broadband collector of NV phonon sideband emission. The proposed
structures can be fabricated using existing materials and processing techniques.
\end{abstract}



\section{Introduction}

\noindent Nitrogen-vacancy (NV) centers found in diamond are a promising system for realizing optically addressable solid-state spin qubits. In
single-crystal diamond, negatively charged NV$^-$ centers have been used in a number of experimental demonstrations relevant to quantum
information processing, including single photon generation \cite{ref:gruber1997sco, ref:kurtsiefer2000sss, ref:beveratos2001nrd}, coherent
population trapping \cite{ref:santori2006cpt}, and optical readout and manipulation of single nuclear spins \cite{ref:jelezko2004oco,
ref:gurudevdutt2007qrb, ref:childress2006cdc}. In order to utilize these properties in quantum information processing applications
\cite{ref:childress2006ftq, ref:benjamin2009pfm}, efficient and scalable optical coupling between NVs and photonic devices such as waveguides
and microcavities \cite{ref:su2008tpt, ref:young2009ces} is necessary. Photonic crystal nanocavities \cite{ref:painter1999tdp,
ref:akahane2003hqp, ref:srinivasan2004ofb, ref:song2005uhq, ref:takahashi2007hqn} confine photons to sub-wavelength mode volumes, $V$, enabling
Purcell enhanced coupling between an optical dipole, such as an NV-center or quantum dot, and the cavity mode \cite{ref:yoshie2004vrs,
ref:hennessy2007qns, ref:englund2007ccr}. The coherent dipole cavity coupling rate scales as $1/\sqrt{V}$, and small $V$ is particularly
beneficial in cavity-QED (Quantum Electrodynamic) systems where intrinsic optical loss or dipole dephasing rates are large.

Recently, optical coupling between high-$Q$ microcavities and NVs hosted in diamond nanocrystals has been reported \cite{ref:park2006cqd,
ref:barclay2008cie, ref:schietinger2008obo}, and devices for studying NVs hosted in nanocrystalline diamond films have been fabricated
\cite{ref:wang2007owg, ref:wang2007fct}. However, to-date many desirable properties of NVs in single-crystal diamond have not yet been observed
in nanocrystalline diamond. Although spontaneous lifetime-limited optical transition linewidths have been measured in single NVs hosted in
diamond nanocrystals, they exhibit $\sim 10$ THz inhomogeneous broadening and have not been measured in high yield from an ensemble of candidate
nanocrystals \cite{ref:shen2008zpl}.  In bulk single crystal diamond, NV$^-$ optical transitions with lifetime limited linewidths
\cite{ref:tamarat2006ssc}, inhomogeneous broadening of less than 10 GHz, and time-averaged spectral diffusion below $100$ MHz
\cite{ref:santori2006cpt} have been observed. Because of the superior properties of NVs in bulk single crystal diamond, some cavity geometries
using only single-crystal diamond and air have been proposed \cite{ref:tomljenovichanic2006dbp, ref:kreuzer2008dpc}. However, there are
difficulties in fabricating photonic devices from single-crystal diamond, since vertical optical confinement within the diamond requires either
a three dimensional etching process, or a method for fabricating thin single-crystal diamond films \cite{ref:fairchild2008fus}.

An alternative to fabricating devices directly from single-crystal diamond is to integrate a patterned high-index optical waveguiding layer on
top of a diamond substrate, from which photons can couple evanescently to NVs close to the diamond surface.  Recently, a hybrid GaP-diamond
material system, consisting of an optically thin GaP film attached to a diamond substrate, was used to demonstrate optical coupling between NVs
close to the diamond surface and ridge waveguides patterned in the GaP film \cite{ref:fu2008cnv}. Owing to the enhanced local density of states
in the near field of the GaP waveguide modes, relatively efficient evanescent coupling between NVs and the GaP waveguide can be achieved.  In
order to efficiently couple to individual NVs, three dimensional confinement provided by an optical cavity is necessary. Here we present designs
for a GaP-diamond photonic crystal nanowire waveguide and nanocavity which can be fabricated using available materials and processing
techniques. Using numerical simulations, we study the sensitivity of the optical properties of these devices on structural parameters. We also
simulate the coupling of spontaneous emission from a broadband source, such as phonon sideband emission from an NV-center, into the proposed
waveguide design, and show that efficient broadband collection of NV emission is possible with these structures.

\section{Hybrid GaP-Diamond photonic crystal nanocavity}

Photonic crystal nanocavities, formed by introducing localized perturbations to planar periodic structures, can support ultra-high $Q/V$
resonances.  Ultra-high $Q/V$ devices have been demonstrated at near-IR wavelengths in free standing membranes such as Si with $Q > 2.5\times
10^6$ \cite{ref:takahashi2007hqn}. In Si films supported by low-index SiO$_2$ substrates (SOI), devices with $Q > 1.5 \times 10^5$ have been
realized \cite{ref:zain2008uhq}.  Although simulations of photonic crystal cavities fabricated from diamond membranes ($n_{\text{Dia}} \sim
2.4$) predict radiation loss limited resonances with $Q > 10^6$ \cite{ref:tomljenovichanic2006dbp, ref:kreuzer2008dpc, ref:bayn2008uhq},
material loss has limited $Q < 600$ \cite{ref:wang2007fct} in nanocavities fabricated from nanocrystalline membranes. Photonic crystal
nanocavities fabricated from single crystal diamond membranes \cite{ref:fairchild2008fus} have not yet been demonstrated.

\begin{figure}[bht]
\begin{center}
  \epsfig{figure=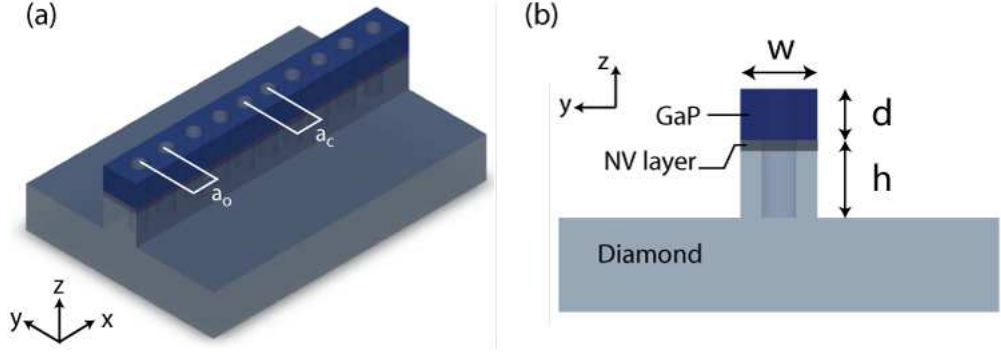, width=1\linewidth}
  \caption{Schematic of the GaP-on-diamond photonic crystal cavity design. (a) Isometric view, (b) end view. For the optimized structure studied in this
  paper, $\left[w,d,h\right] = \left[192,128,640\right]$ nm, $\left[a_o,a_c\right] = \left[160,141\right]$ nm (tapered over 6 periods),
  and the hole radius $r = 43$ nm.}\label{fig:pc_schematic}
\end{center}
\end{figure}

Here we study photonic crystal cavities formed in a thin waveguiding layer supported by a diamond substrate hosting high quality NVs near the
diamond surface, as indicated in Fig. \ref{fig:pc_schematic}.  In choosing a film from which to form the waveguide layer, we are limited to
materials which are transparent at the NV$^-$ zero phonon transition wavelength, $\lambda_{\text{NV}^-} = 637$nm, and whose refractive index
exceeds $n_\text{Dia}$. Epitaxially grown GaP films have a nominally high refractive index ($n_{\text{GaP}}\sim 3.3 > n_\text{Dia})$ and low
optical absorption at $\lambda_{\text{NV}^-}$. Photonic crystal cavities formed in GaP membranes have recently been demonstrated with $Q \sim
1700$ \cite{ref:rivoire2008gpp}, and in Ref.\ \cite{ref:fu2008cnv}, optical loss of $\sim 72$ dB/cm was measured in a GaP waveguide supported by
a single crystal diamond substrate.  The waveguide loss measurements presented in Ref.\ \cite{ref:fu2008cnv} indicate that devices with
absorption and surface scattering limited $Q > 2 \times 10^4$ can be realized in this system. With improved processing to reduce scattering
loss, it is expected that material loss limited $Q$ exceeding this value should be possible.

Realizing a high-$Q$ photonic crystal nanocavity in this GaP-diamond system is complicated by the diamond substrate's moderately high refractive
index ($n_\text{Dia} \sim 2.4$). Compared to photonic crystal membrane or SOI devices, a diamond substrate expands the light cone into which
photons can radiate out of the nanocavity.  In addition, the broken vertical mirror symmetry from the diamond substrate precludes the existence
of a bound mode in the GaP waveguide layer for arbitrarily low frequencies. As discussed below, the underlying GaP-diamond nanowire waveguide
structure from which the nanocavity studied here is realized does not support a non-radiating mode in the wavelength range of interest. This is
a result of the subwavelength dimensions of the waveguide cross-section, which provides strong modal confinement at the expense of reducing the
effective index of the waveguide modes below $n_{\text{Dia}}$. However, by extending the waveguide sidewalls into the diamond substrate,
radiation into the diamond substrate can be made arbitrarily small.  As we will show below, for a realistic diamond sidewall height, the
proposed structure supports waveguide modes whose radiation loss is smaller than the expected intrinsic material or scattering loss, and forms a
low-loss structure for realizing a high-$Q$ nanocavity.

\subsection{GaP-diamond photonic crystal nanowire waveguide}\label{sec:waveguide}

The underlying structure of the photonic crystal nanocavity illustrated in Fig.\ \ref{fig:pc_schematic} consists of a GaP photonic crystal
nanowire waveguide whose sidewalls have been extended into a diamond substrate. This waveguide supports Bloch modes which form the basis for
localized resonances formed when the periodic symmetry of the waveguide is broken. In general, the Bloch modes are either leaky or guided,
depending on their frequency, $\omega$, and wavenumber, $k$, relative to the air and diamond lightlines \cite{ref:johnson2000lwp,
ref:lalanne2002eap}. Waveguide modes with an effective index smaller than that of diamond, $n_{\text{eff}} = k/\omega(k) < n_{\text{Dia}}$
(setting $c = 1$), will leak into the diamond substrate. High-$Q$ cavity resonances are realized by engineering a perturbation which couples
guided or low-loss Bloch modes \cite{ref:srinivasan2002msd, ref:painter2003lds}. Using 3-dimensional finite difference time domain (FDTD)
simulations \cite{ref:meep} to study the photonic crystal waveguide loss, we can predict an upper limit on the $Q$ of a nanocavity formed from
this structure.

\begin{figure}[bt]
\begin{center}
  \epsfig{figure=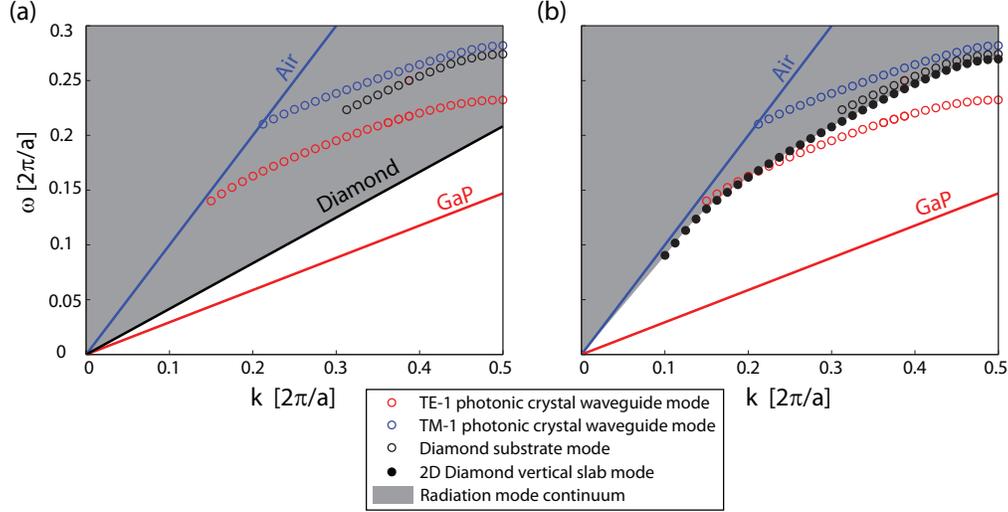, width=1\linewidth}
  \caption{GaP-diamond photonic crystal waveguide band structure, and GaP, diamond substrate, and air lightlines.  Shaded regions indicate the presence
  of a continuum of lossy radiating modes. In both (a) and (b) the photonic crystal waveguide modes (red and blue points) were calculated for a
  structure with  $h = 640$ nm.  The diamond lightline in (a) is that of a bulk diamond substrate. The filled black points in (b) are the lowest
  frequency $E_y$ polarized modes of an infinitely tall diamond slab of width $w$ patterned with air holes along the vertical axis whose spacing and
  radius are equal to that of the GaP waveguiding layer.
 }\label{fig:pc_wg_band}
\end{center}
\end{figure}

Figure \ref{fig:pc_wg_band}(a) shows the first Brillouin zone of the band-structure for the lowest frequency GaP-diamond photonic crystal
nanowire waveguide modes. Also shown are the air, diamond, and GaP lightlines. The calculation used waveguide cross-section dimension
$\left[w,d\right] = \left[192,128\right]$ nm, hole spacing $a = 160$ nm, and hole radius $r = 43$ nm. The hole spacing was chosen to
approximately position the lowest energy waveguide mode band edge in the wavelength range of interest. The waveguide cross-section was not
systematically optimized, and when normalized by the hole spacing $a$, is similar to that used in previous nanowire based cavity designs
\cite{ref:zain2008uhq, ref:velha2006uhr}. $r/a$ was chosen based upon initial simulations of the $Q$ of the nanocavity design studied below in
Sec.\ \ref{sec:defect}, and is also similar to values used in other work. A mesh with resolution $a/20$, perfectly matched layer (PML) boundary
conditions in the $\hat{y}$ and $\hat{z}$ directions ($320$ nm PML layer thickness), and Bloch boundary conditions in the $\hat{x}$ direction
were used in these simulations. For clarity, only modes with even parity in the $\hat{y}$ direction are calculated. Odd modes lie higher in
frequency, and as will become clear below, we are primarily interested in the lowest frequency waveguide mode.  Note that parity in the
$\hat{z}$ dimension is not conserved for these GaP-diamond structures, since they are not vertically symmetric.  Both the lowest frequency
TE-like (dominantly $E_y$ polarized) and TM-like (dominantly $H_y$ polarized) modes are shown in Fig.\ \ref{fig:pc_wg_band}. We label these
modes TE-1 and TM-1, respectively. Also evident in Fig.\ \ref{fig:pc_wg_band}(b) is a mode (black open points) whose dispersion lies between the
TE-1 and TM-1 bands. This mode corresponds to a leaky mode whose field is predominantly confined within the diamond substrate. Leaky modes with
waveguide quality factor, $Q_{\text{wg}} < 100$ are not shown. $Q_\text{wg}$ is related to the waveguide loss per unit length, $\alpha$, by
$\alpha = \omega/ Q v_g$, where $v_g = \partial \omega /\partial k$ is the group velocity of the waveguide mode.

From Fig.\ \ref{fig:pc_wg_band}(a), it is immediately clear that for the specified waveguide dimensions, no waveguide modes lie below the
diamond lightline.  In particular, at the band edge ($k=k_{X}=\pi/a$) the effective index of the lowest frequency waveguide mode is smaller than
$n_\text{Dia}$; this mode is expected to leak into the diamond substrate.   This leakage could be reduced by increasing the GaP waveguide
dimensions ($w,d$), increasing the effective index of the mode at the expense of decreasing the peak single photon field strength, and
decreasing the frequency and momentum mismatch between the TE-1 and higher order waveguide modes.  Alternatively, we can extend the GaP
sidewalls (including the holes) by depth $h$ into the diamond substrate. In the limit that $h\gg\lambda/n_\text{Dia}$, the bulk diamond
lightline is no longer relevant in determining whether a GaP-diamond waveguide mode is bound or leaky.  Instead, the relevant substrate is the
approximately two dimensional vertical diamond photonic crystal slab underneath the GaP photonic crystal waveguide.  The band-structure of the
lowest frequency $E_y$ polarized mode supported by this vertical slab is shown in Fig.\ \ref{fig:pc_wg_band}(b).  Above this band exists a
continuum of radiating slab modes with non-zero vertical momentum; this band forms a renormalized ``structured lightline''. At $k=k_{X}$, the
TE-1 mode lies below this structured lightline, indicating that this mode is not leaky in the $h\gg\lambda/n_\text{Dia}$ limit. The existence of
this mode below the structured lightline is not sensitive to the choice of $w$, since decreasing $w$ decreases the effective index of both the
TE-1 mode and the lowest energy radiating slab mode.  Decreasing $d$ decreases the TE-1 effective index in a manner typical of slab and ridge
waveguides \cite{ref:snyder1983owt}, and can push the TE-1 mode above the structured lightline at a given frequency.

\begin{figure}[bt]
\begin{center}
  \epsfig{figure=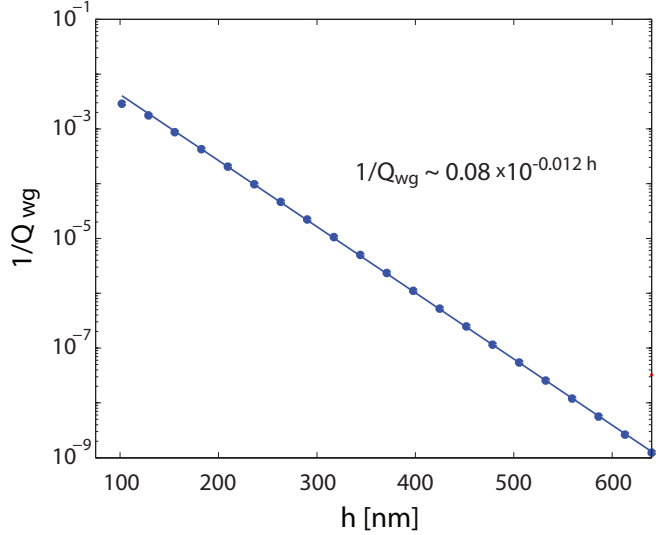, width=0.65\linewidth}
  \caption{Dependence of the photon loss rate ($\sim \omega / Q_\text{wg}$) of the TE-1 photonic crystal waveguide mode on etch depth, for $k=\pi/a_c$,
  and hole spacing $a=a_c = 141$ nm).}\label{fig:pc_wg_Q}
\end{center}
\end{figure}

In Fig.\ \ref{fig:pc_wg_Q} the waveguide loss per optical cycle, $1/Q_\text{wg}$, of the TE-1 mode is shown as a function of the etch depth $h$,
at $k = k_X$, for $a = a_c = 141$ nm, where $a_c$ is the ``cavity'' hole spacing of the nanocavity design presented in the following section and
indicated in Fig.\ \ref{fig:pc_schematic}.  This was calculated using FDTD simulations, by monitoring the power radiated into the absorbing
$\hat{y}$ and $\hat{z}$ boundaries of a waveguide Bloch unit cell. We expect $Q_\text{wg}$ at $k=k_X$ to place an upper limit on the $Q$ of a
cavity mode formed predominantly from the TE-1 mode \cite{ref:painter2003lds}. Figure \ref{fig:pc_wg_Q} indicates that for $h<100$ nm, the TE-1
waveguide mode is very leaky with $Q_\text{wg} < 500$, but that for increased etch depth $h = 600$ nm, the loss can be reduced and $Q_\text{wg}
> 10^9$. For $h > 100$ nm, $Q_\text{wg}$ increases exponentially with $h$.

\subsection{Nanocavity design}\label{sec:defect}

\begin{figure}[bt]
\begin{center}
  \epsfig{figure=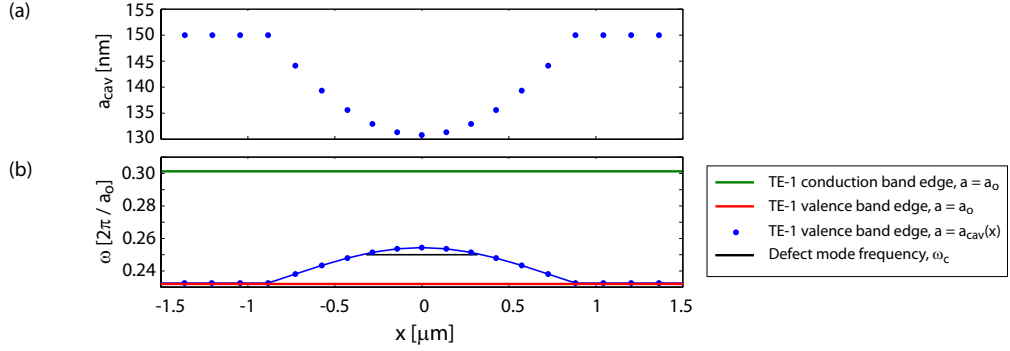, width=1\linewidth}
  \caption{(a) Lattice constant $a_{cav}$ as a function of position in the cavity.  Each point, $x_i$, on the graph corresponds to a
  position midway between the center of two holes spaced by $a_{cav}(x_i)$. (b) Frequency of the photonic crystal waveguide TE-1 valence band edge,
  $\omega^\text{TE-1}_X = \omega^\text{TE-1}(k_X=\pi/a)$, for $a$ set by the cavity's varying ``local'' hole spacing, $a_{cav}(x_i)$, shown in (a).}\label{fig:pc_heterostructure}
\end{center}
\end{figure}

\begin{figure}[bt]
\begin{center}
  \epsfig{figure=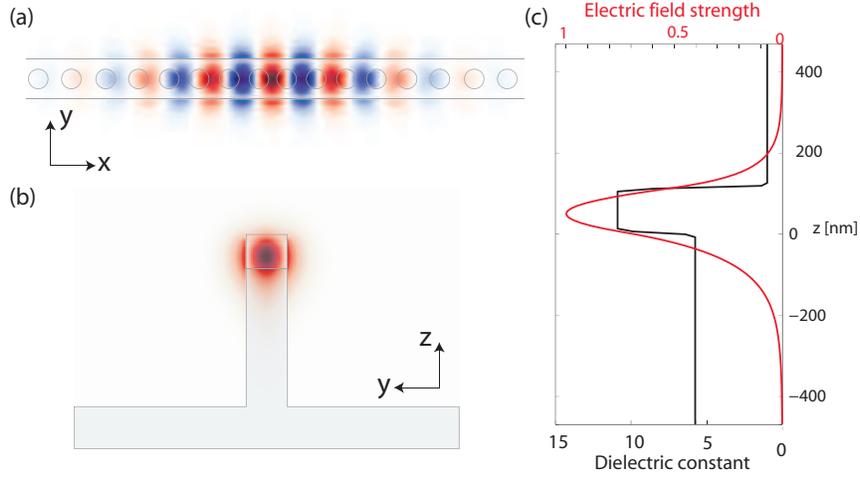, width=0.85\linewidth}
  \caption{Dominant electric field component ($E_y$) of the high-$Q$ nanocavity mode.  (a) Top view: $x-y$ plane bisects the GaP waveguiding layer. (b)
  End view: $y-z$ plane through the center of the cavity. (c) Vertical profile of the field and dielectric constant.}\label{fig:pc_cavity_field}
\end{center}
\end{figure}

The nanocavity studied here is formed in the photonic crystal nanowire waveguide described in Sec.\ \ref{sec:waveguide} by locally perturbing
the hole spacing, $a_{cav}$, as indicated in Figs.\ \ref{fig:pc_heterostructure}(a) and \ref{fig:pc_cavity_field}(a), forming a heterostructure
\cite{ref:istrate2006pch} cavity \cite{ref:song2005uhq, ref:notomi2008uqn}. The hole spacing is varied slowly over 6 periods from $a_c = 141$ nm
in the center of the cavity, to a constant ``bulk'' value of $a_o=160$ nm. The cavity is symmetric about the $x$ and $y$ axes. For simplicity,
the hole radius remains constant.  This type of photonic crystal nanowire waveguide cavity \cite{ref:foresi1997pbm, ref:jugessur2003odp}
supports localized cavity modes with simulated radiation loss limited $Q > 10^7$ when applied to structures with air undercladding
(waveguide-substrate index contrast $\Delta n = n_\text{SiN} - n_\text{air} \sim 2.0 - 1.0$) \cite{ref:mccutcheon2008duq, ref:chan2009omd}. In
SOI material systems ($\Delta n = n_\text{Si} - n_{\text{SiO}_2} \sim 3.48 - 1.44$), nanowire based cavities with $Q> 4\times10^5$ have been
theoretically studied \cite{ref:velha2006uhr}. Ultrahigh $Q/V$ nanocavities formed in photonic crystal nanowire waveguides on a high-index
substrate such as diamond have not been previously investigated.

Adjusting the local hole spacing within the photonic crystal waveguide shifts the frequency range of the photonic crystal waveguide stop-band.
In a quasi-1D picture, the photonic crystal waveguide forms an ``optical potential'' \cite{ref:painter2003wle} whose band edge is locally
modulated by the variation in the hole spacing or hole size \cite{ref:barclay2003dpc, ref:morie2004dco, ref:chan2009omd}. In the cavity design
considered here, the local reduction in hole spacing shifts the valence and conduction band edges up in frequency so that the cavity supports an
``acceptor'' defect mode \cite{ref:johnson2000lwp} formed by superpositions of the valence band states of the unperturbed photonic crystal
nanowire waveguide \cite{ref:painter2003lds}.

Following an analysis similar to that in Ref.\ \cite{ref:chan2009omd}, the photonic crystal cavity optical potential is illustrated by Fig.\
\ref{fig:pc_heterostructure}(b), which shows the frequency $\omega_X^\text{TE-1}$ of the TE-1 valence band edge as a function of the ``local''
hole spacing, $a_{cav}$, at positions $x_i$ midway between two holes in the cavity. The band edge frequency $\omega_X^\text{TE-1}$ was
calculated at each $x_i$ by using FDTD simulations to determine the band structure of the nanowire photonic crystal waveguide in Sec.\
\ref{sec:waveguide}, with the hole spacing set to $a_{cav}(x_i)$. Also shown are the frequencies of the TE-1 valence and conduction band edges
in the ``bulk'' waveguide region, where $a=a_o$, and of a high-$Q$ localized cavity mode discussed in detail below. Note that these photonic
crystal nanowire cavities have also been analyzed in the context of the unperturbed photonic crystal waveguide sections forming mirrors for
photons trapped in the central cavity region \cite{ref:lalanne2004tpm, ref:sauvan2005mre, ref:velha2006uhr, ref:mccutcheon2008duq}.

\begin{figure}[bt]
\begin{center}
  \epsfig{figure=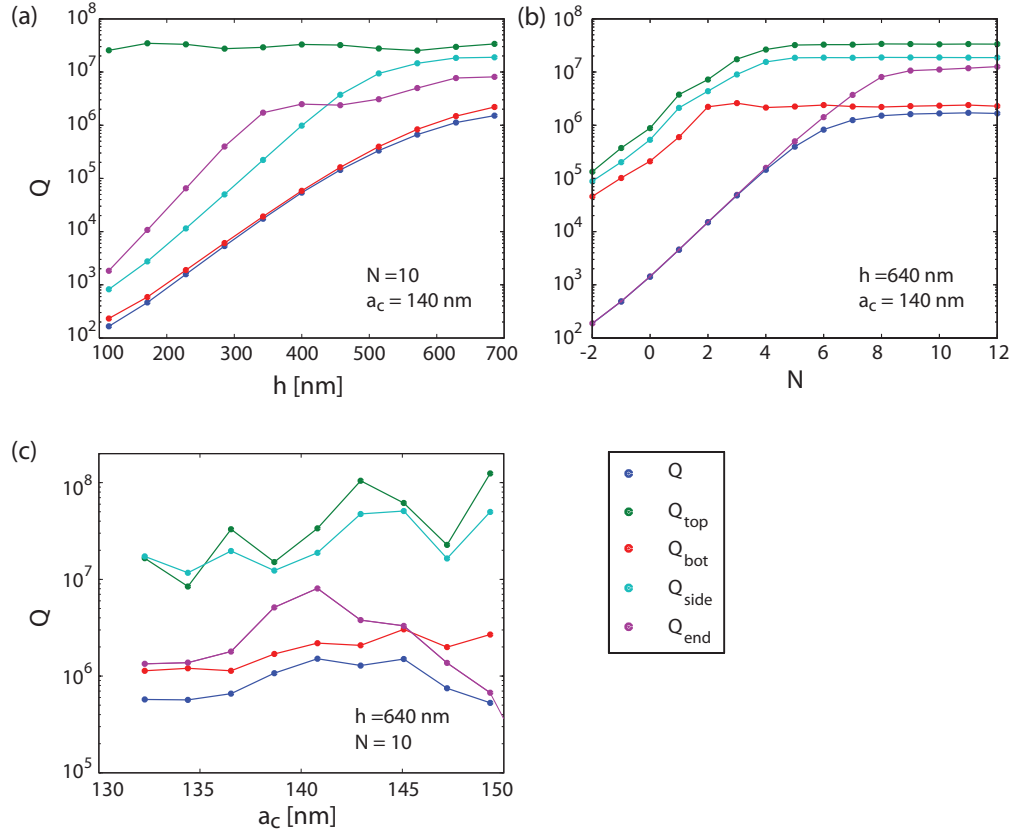, width=1.0\linewidth}
  \caption{Dependence of nanocavity mode $Q$ on geometric parameters.  The total $Q$ (blue points) and the contribution to $Q$ from radiation into
  specific directions (other colored points) are given.  Dependence of $Q$ on (a) etch depth $h$, (b) number of ``mirror'' periods
  $N$ in the $\hat{x}$ direction between the edge of the graded cavity region and the PML absorbing simulation boundary,
  (c) cavity minimum (center) hole spacing $a_c$.  For $N<0$ in (b), the PML boundary of the simulation domain overlaps $|N|$ periods of the
  graded cavity region.
}\label{fig:pc_cav_Q_vs_p}
\end{center}
\end{figure}

FDTD simulations of this structure predict that it supports a mode at a resonance wavelength $\omega_c$ close to the NV$^-$ zero phonon line at
637nm ($\omega_c = 0.249 \times 2\pi/a_o$), with $Q = 1.5 \times 10^6$ and mode volume $V = 0.52 \left(\lambda/n_{\text{GaP}}\right)^3$, where
$V$ is defined by the peak electric field energy density: $V = \int n^2 |E|^2 d\vrr / \left(n^2|E|^2\right)_\text{max}$.  The maximum field
amplitude inside the diamond is $E_s = 0.72 E_o$, where $E_o$ is the field maximum at the location of maximum energy density.  $E_o$ is located
inside the high index GaP layer for the device and mode studied here. The field maximum decays to $0.10 E_o$ at a depth $z = 155$ nm below the
diamond surface. Simulations indicate that similar designs with the number of grading periods reduced to 4 support more tightly confined modes
($V \sim 0.3 \left(\lambda/n_{\text{GaP}}\right)^3$) with a maximum $Q \sim 6 \times 10^4$. The mesh used in these simulations had uniform
resolution $a_o/20$, and extended $N=8$ periods ($a_o$) in $\pm \hat{x}$ beyond the end of the graded region, $5 a_o$ in $\pm\hat{z}$  above and
below the diamond surface, and $3 a_o$ along the $\pm \hat{y}$ directions.  A PML layer thickness of $4 a_o$ was used in the highest-$Q$
simulations. After exciting the structure with a narrow band source, the steady state ratio of stored and radiated power was used to calculate
the modal $Q$. The radiated power along each axis was monitored using reference frames positioned $a_o/2$ away from each PML boundary.  Mirror
symmetry was enforced about the $\hat{x}$ (odd parity) and $\hat{y}$ (even parity) axes.  Higher resolution ($a_o/25$) simulations of the
$h=640$ nm structures were also conducted, and it was verified that the structure supports a mode with $Q > 1.5\times10^6$.

The role of the diamond substrate in limiting $Q$ is illustrated by simulating its dependence on the depth of the etched diamond ridge. Figure
\ref{fig:pc_cav_Q_vs_p}(a) shows $Q$ as a function of $h$, and indicates the relative contributions to $Q$ due to radiation into the bottom
diamond substrate ($Q_\text{bot}$), top air cladding ($Q_\text{top}$), end of the cavity ($Q_\text{end}$) and side of the cavity
($Q_\text{side}$).  The total cavity $Q$ is given by $Q^{-1} = Q_\text{top}^{-1} + Q_\text{bot}^{-1} + Q_\text{end}^{-1} + Q_\text{side}^{-1}$.
 When $h$ is reduced to $160$ nm, increased radiation loss into the substrate degrades $Q < 500$.  For $h > 640$ nm, radiation loss through the
end of the cavity into the photonic crystal waveguide begins to play a non-negligible role in limiting $Q$.

Figure \ref{fig:pc_cav_Q_vs_p}(b) shows the dependence of $Q$ on the number $N$ of waveguide periods between the cavity grading edge and the PML
absorbing layer used in the simulation.  For $N>8$, $Q_\text{end}$ is approximately constant, indicating that it is limited by coupling between
the cavity mode and either photonic crystal waveguide modes which do not exhibit a bandgap at $\omega_c$ (e.g., the TM-1 mode) or radiation
modes \cite{ref:lalanne2004tpm, ref:sauvan2005mre}. For $N < 5$, the cavity loss is dominated by ``mirror'' leakage. In a practical device, $N$
represents the number of ``mirror'' periods between the edge of the cavity grading and an integrated waveguide, and would be set to a value that
ensures that radiation loss from of the cavity is predominantly into the waveguide.

The sensitivity of $Q$ to the central cavity hole spacing, $a_c$, is shown in Fig.\ \ref{fig:pc_cav_Q_vs_p}(c).  When $a_c$ is varied between
$132 - 150$ nm, $Q$ remains $> 5 \times 10^5$.  In these simulations, the bulk hole spacing is maintained fixed at $a_o = 160$ nm, and the
grading of $a_c$ remains parabolic.  $Q$ and $Q_\text{end}$ are maximized when $a_c = 141$ nm, while $Q_\text{bot}$ trends upward as $a_c \to
a_o$. In this limit, the depth of the parabolic cavity defect becomes increasingly small, minimizing coupling to lossy waveguide modes close to
the diamond lightline.  As a result, $Q_\text{bot}$ increases. $Q_\text{end}$ is maximized when a sufficiently small $a_c$ is chosen to position
$\omega_c$ well within the TE-1 bandgap, as illustrated in Fig.\ \ref{fig:pc_heterostructure}(b). However, pushing $\omega_c$ too far above the
TE-1 valence band edge can result in efficient phase matched coupling between the cavity mode and the lossy TM-1 valence band edge modes (see
Fig.\ \ref{fig:pc_wg_band}), reducing $Q_\text{end}$. These competing dependencies determine the optimum $a_c$. Note that $V$ varies between
$0.42-0.70$ $(\lambda/n_\text{GaP})^3$ over the simulated range of $a_c$.

\subsection{Nanocavity cavity-QED parameters for coupling to NV$^-$ centers}

From the FDTD calculations of $V$ and $Q$ presented above, we can predict the coherent coupling rate, $g$, for a single NV$^-$ center located
near the cavity and coupled to a single photon stored in the nanocavity mode. For experiments in cavity QED we must be careful to distinguish
between zero-phonon optical transitions (these become spectrally narrow at low temperature) and phonon-assisted transitions.  The total
spontaneous emission rate of an NV$^-$ center is measured to be $\gamma_\text{tot} = 2\pi\times13$ MHz; however the rate into the ZPL alone,
$\gamma_\text{ZPL}$, is only $3\%$ of $\gamma_\text{tot}$ \cite{ref:manson06nit}. The coherent coupling rate between a single photon and the
NV$^{-}$ ZPL is given by
\begin{equation}\label{eqn:g}
g_\text{NV}/2\pi =
  \frac{1}{8\pi^2}\sqrt{\frac{3\omega\gamma_\text{ZPL}}{\overline{V}}\frac{n_\text{GaP}}{n_\text{Dia}}}\left|\frac{E(\vrr_\text{NV})}{E_o}\right|
\end{equation}
where $\overline{V}=V/(\lambda/n_\text{GaP})^3$, $|E(\vrr_\text{NV})|$ is the magnitude of the electric field at the NV location, and $E_o$ is
the field strength at the electric field energy density maximum, which is located within the GaP for the device studied here. This expression
assumes that the spatial orientation of the NV transition dipole moment and electric field polarization are parallel, which is possible for one
of four allowed NV orientations in $\langle 111 \rangle$ diamond. For the cavity mode volume given in Sec.\ \ref{sec:defect}, Eq.\ (\ref{eqn:g})
gives $g_s/2\pi =  2.25$ GHz for an NV optimally located at the diamond surface.

Assuming that the cavity $Q$ is limited by radiation loss as calculated in Sec.\ \ref{sec:defect}, we see that
$\left[g_s,\kappa,\gamma_\text{tot}\right]/2\pi = \left[2.25,0.16,0.013\right]$ GHz, where $\kappa = \omega/2Q$, indicating that it is possible
to reach the strong coupling regime \cite{ref:kimble1998sis} using this nanocavity design. Also of interest is the Purcell enhanced spontaneous
emission rate of the NV$^-$ ZPL into the cavity mode. Assuming the cavity and NV$^-$ ZPL are aligned spectrally, the Purcell enhancement factor
to the ZPL spontaneous emission rate is given by,
\begin{equation}\label{eqn:Fp}
F = \frac{3}{4\pi^2}\frac{Q}{\overline{V}}\frac{n_\text{GaP}}{n_\text{Dia}}\left|\frac{E(\vrr_\text{NV})}{E_o}\right|^2\frac{\gamma_\text{ZPL}}{\gamma_\text{tot}}
 = \frac{2 g^2_\text{NV}}{\kappa\gamma_\text{tot}}.
\end{equation}
For an NV optimally positioned at the diamond surface, Eq.\ (\ref{eqn:Fp}) gives $F_s = 4.9\times10^3$.

In a realistic experiment, it may not be possible to couple to NVs with desirable optical and spin coherence properties arbitrarily close to the
diamond surface \cite{ref:santori2008vdn}.  In addition, material loss and fabrication imperfections will likely limit the nanocavity $Q$ below
the highest values presented in Sec.\ \ref{sec:defect}.  If we assume that the NV is optimally positioned $50$ nm below the diamond surface, and
that the cavity $Q$ is limited to $2 \times 10^4$ (e.g., due to material absorption and surface scattering \cite{ref:fu2008cnv}), we find $F
> 16$, indicating that $\sim 94\%$ of the \emph{total} NV$^-$ spontaneous emission will radiate into the nanocavity mode.  The $6\%$ of
uncoupled NV emission will radiate into phonon sidebands not resonant with the cavity mode.

\subsection{Waveguide collection of NV phonon sideband emission}

In addition to forming the basic structure for the the nanocavity presented in Sec.\ \ref{sec:defect}, the waveguide studied in Sec.\
\ref{sec:waveguide} can function as an efficient broadband collector of radiation from NV phonon sidebands. Measurements of broadband
phonon-sideband emission are used in experiments to gather information regarding the relative populations of the NV spin triplet ground states
\cite{ref:santori2006cpt}, as well as the nuclear spin state of neighboring impurities \cite{ref:jelezko2004oco, ref:gurudevdutt2007qrb,
ref:childress2006cdc}. From coupled-mode theory \cite{ref:snyder1983owt}, the normalized power spectral density, $\left|s(\omega)\right|^2$,
coupled into a waveguide mode from a NV dipole located at position $\vrr_\text{NV}$ in the photonic crystal waveguide near field is
approximately given by
\begin{equation}\label{eqn:a_wg}
\left|s(\omega)\right|^2 = \frac{3}{8\pi}\left|\frac{E(\vrr_\text{NV})}{E_{o}}\right|^2\frac{(\lambda/n_\text{GaP})^2}{A}\frac{n_g(\omega)}{n_\text{Dia}},
\end{equation}
where $n_g(\omega)$ is the frequency dependent group index of the waveguide mode, $E(\vrr)$ is the waveguide mode field amplitude, and $A$ is
the waveguide mode area defined by
\begin{equation}
A = \frac{1}{a_o}\frac{\int_u d\vrr \, n^2(\vrr) \left|E(\vrr)\right|^2}{\left(n(\vrr)^2 \left|E(\vrr)\right|^2\right)_\text{max}},
\end{equation}
where volume $u$ is a unit cell of the waveguide. We have assumed that the dipole and waveguide field at $\vrr_\text{NV}$ are parallel.
$|s(\omega)|^2$ is normalized by the ``bulk'' power spectral density of the NV dipole embedded far below an unpatterned diamond surface.
Equation (\ref{eqn:a_wg}) indicates that coupling into a given waveguide mode can be increased by maximizing $n_g/A$. Photonic crystal
waveguides are ideally suited in this regard \cite{ref:mangarao2007sqd}: they can support modes with sub-wavelength modal area and large $n_g$.

\begin{figure}[bt]
\begin{center}
  \epsfig{figure=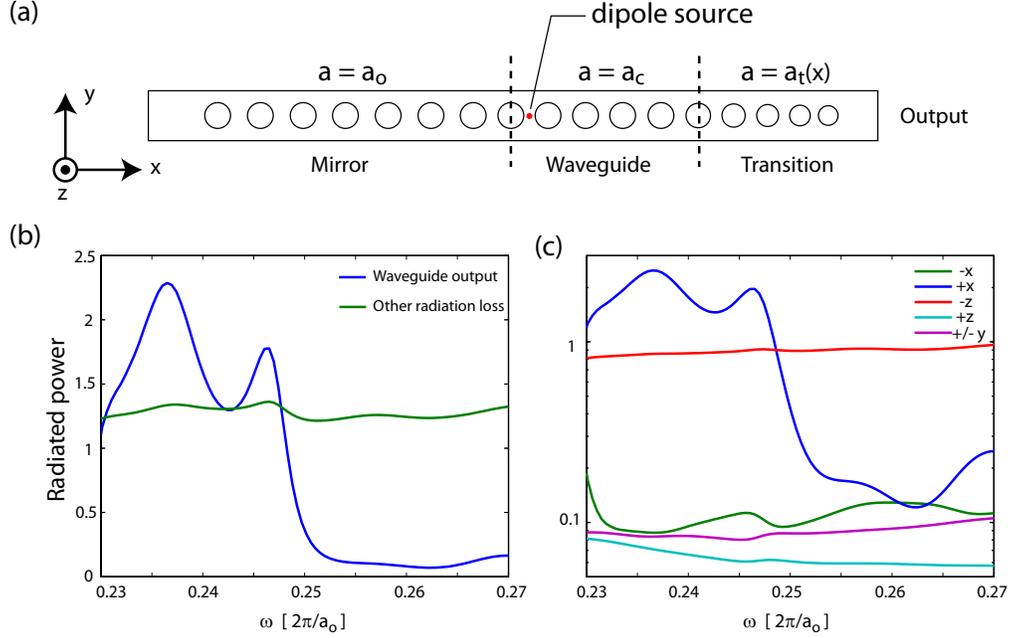, width=1\linewidth}
  \caption{(a) Geometry of the simulated dipole and the asymmetric photonic crystal waveguide hole pattern.  Left of the dipole, the hole spacing is
  set to a value ($a_o$) larger than the hole spacing ($a_c$) to the right of the dipole.  For frequencies close to the valence band edge $\omega_X^\text{TE-1}$
  of the right waveguide region, there are no propagating waveguide modes to the left of the dipole. In the transition region, the hole spacing and hole radius are graded to $0.75$ of their nominal values in order to reduce back reflections
  \cite{ref:sauvan2005mre}.  More precisely, the hole spacings and hole radii in the transition region are scaled by $\left[0.95, 0.86, 0.8, 0.75\right]$
  (moving in the $+\hat{x}$ direction) from the values in the waveguide region.
  (b) Normalized radiated spectra of the dipole into waveguide and non-waveguide radiation channels.
  (c) Normalized radiated spectra of the dipole into the principal directions of the simulation domain.
}\label{fig:spectrum}
\end{center}
\end{figure}

To gain a quantitative estimate of the NV-waveguide coupling, and to study enhancement or suppression of NV coupling into other radiation modes
of the waveguide structure, it is useful to conduct FDTD simulations of the emission spectrum of a broadband dipole source positioned at
$\vrr_\text{NV}$. Here we consider a broadband dipole, polarized along $\hat{y}$ and located $z_\text{NV} = 11$ nm below the diamond surface and
midway between two holes of the photonic crystal waveguide studied in Sec.\ \ref{sec:waveguide}.  In order to channel the dipole emission into a
forward propagating waveguide mode, the hole spacing of the waveguide structure used here is asymmetric about the dipole position, as shown in
Fig.\ \ref{fig:spectrum}(a). For $a_o > a_c$, the valence band edge of this structure is lower on the left side of the dipole than on the right
side; the band edge of the left (``mirror'') region has a lower frequency than the band edge of the right (``waveguide'') region. For efficient
collection of NV emission, the structure should be designed such that the majority of the NV phonon sideband emission is lower in frequency than
the ``waveguide'' TE-1 valence band edge, and higher in frequency than the ``mirror'' TE-1 valence band edge. Note that reflections from the
waveguide termination can dramatically alter the coupling between a dipole and the waveguide, as studied in \cite{ref:mangarao2007sqd}. Here we
have chosen a termination designed to suppress these reflections in the simulations. In FDTD simulations of periodic structures supporting
propagating modes, PML terminations of the propagating axis can create significant reflections.  A waveguide which is invariant along the
propagating dimension, such as the termination region of the structure considered here, is better suited for a PML boundary.

Figure \ref{fig:spectrum}(b) shows the FDTD calculated spectrum for the system described above. Power spectral density of radiation into the
waveguide, and into all other channels (e.g., the substrate) are shown.  In the simulations studied here, $a_c = 141$ nm and $a_o = 160$ nm. The
hole radius and waveguide cross-section are the same as in Sec.\ \ref{sec:waveguide}, and $h = 4a_o$.  The spectra are normalized by the
spectrum of the identical dipole source when it is positioned at depth $z_\text{NV}$ below an unpatterned diamond surface. From Fig.\
\ref{fig:spectrum}(b), we see that for frequencies below the valence band edge of the waveguide region ($\omega < \omega_X^\text{TE-1} \sim 0.25
\times 2\pi/a_o$), the emission into the waveguide exceeds the total ``bulk'' emission in absence of the waveguide. The radiated waveguide power
was calculated by monitoring the power flux through an area overlapping the waveguide cross-section, with dimensions $2w\times 2d$, located near
the simulation PML boundary. Note that emission into non-waveguide modes is also enhanced.  As shown in Fig.\ \ref{fig:spectrum}(c), radiation
into the substrate ($-\hat{z}$ direction) and into the waveguide ($+\hat{x}$ direction) are the dominant radiation channels.  Also, note that
relatively little power is coupled into the backward propagating waveguide mode ($-\hat{x}$ direction) in the displayed frequency range. For
$a_o = 160$ nm, the bandwidth of this efficient waveguide coupling is approximately $637 - 700$ nm, allowing efficient collection of a large
portion of the NV$^-$ zero phonon line and phonon sideband emission. This may be particularly useful in room-temperature NV experiments which do
not rely upon readout of a sharp optical ZPL, but require efficient collection of sideband emission, for electron spin readout, for example.

\subsection{Fabrication considerations}
Fabrication of the devices studied in this paper requires three key processing abilities: (i) patterning GaP or another high index film with the
design presented above, (ii) attaching high index films to a diamond substrate, and (iii) transferring the thin film pattern into the diamond
substrate. Proofs of principle of these individual steps have been demonstrated.  Results presented in Refs.\ \cite{ref:rivoire2008gpp,
ref:wang2007fct} show that it is possible to lithographically define and etch sufficiently small holes using current electron beamwriting and
plasma etching tools. Large area patterned GaP films were transferred and adhered \cite{ref:yablonovitch1990vdw} onto a diamond sample in Ref.\
\cite{ref:fu2008cnv}, as have smaller micron-scale structures \cite{ref:barclay2009mfc}. Transferring the pattern into the diamond can be
accomplished using oxygen based plasma etching, e.g., in Ref.\ \cite{ref:santori2008vdn} a highly selective plasma etch was used to extend a SiN
pattern into a diamond substrate.  The degree to which the precise design studied in Sec.\ \ref{sec:defect} can be replicated will affect the
maximum achievable $Q$, and achieving the high aspect ratio diamond sidewalls of the $h=640$ nm structure may be challenging. Theoretical
studies analyzing the importance of the sidewall slope may be necessary in future work. Finally, note that although precise (2 nm) tuning of
$a_c$ was considered in Sec.\ \ref{sec:defect} in order to maximize $Q$, the cavity supports modes with $Q > 5 \times 10^5$ over the range 132
nm $\le a_c \le $ 150 nm.


The GaP-based cavity proposed here is designed specifically for coupling to negatively charged NV centers positioned within 50 nm of a diamond
surface.  It is well established that a dense layer of NV centers close to a diamond surface can be created in high-nitrogen diamond by ion
implantation and annealing~ \cite{ref:davies1992vrc, ref:meijer2005gsc, ref:waldermann2007cdc, ref:santori2008vdn}.  Such a dense layer is
well-suited for initial testing of the optical characteristics of GaP-based microcavities.  However, for the proposed devices to be useful for
quantum information applications, single NV centers with good spectral characteristics must be fabricated close to a surface in high-purity
diamond. Initial tests suggest that in high-purity diamond, a nearby surface can have a large effect on NV properties~\cite{ref:santori2008vdn}
that include charge stability and optical linewidth. A thorough study of how to optimize the properties of single NV centers close to a surface
will be required for any device based on NV centers coupled to an optical microcavity, including the GaP-based design presented here.

\section{Conclusion}

In this paper we have analyzed a photonic crystal geometry which supports low-loss, subwavelength nanocavities and waveguides suitable for
coupling to diamond NV-centers in single crystal diamond.  The structures presented here can be fabricated using existing materials and
processing techniques, and promise to allow efficient collection of photons emitted from NV-centers, a crucial requirement for proposed
applications \cite{ref:benjamin2009pfm}.  Efficient collection of  NV ZPL emission is necessary for single photon and quantum repeater
applications of NVs, and efficient collection of NV sidebands is required for high bandwidth readout of the NV electron spin.  The photonic
crystal nanocavity design presented above supports modes with the necessary $Q$ and $V$ to reach the strong coupling or large Purcell factor
regimes with the ZPL of an NV-center.  The photonic crystal waveguide analyzed here allows broadband collection of NV emission, and is promising
for efficient collection of NV sidebands.  In future work, it is expected that integration of multiple cavities, and realization of nanophotonic
circuits for quantum information processing using NV-centers, will be possible using the device designs presented here.

\end{document}